\documentclass[%
 reprint,
 amsmath,amssymb,
 aps,
]{revtex4-1}

\usepackage{graphicx}
\usepackage{dcolumn}
\usepackage{bm}
\usepackage[
      colorlinks=true,
      linkcolor=blue,
      urlcolor=blue,
      filecolor=black,
      citecolor=blue,
            ]{hyperref}

\begin{document}

\preprint{APS/123-QED}

\title{Collapse of self-interacting scalar field in anti-de Sitter space}

\author{Rong-Gen Cai}
 \email{cairg@itp.ac.cn}
\author{Li-Wei Ji}
 \email{jiliwei@itp.ac.cn}
\author{Run-Qiu Yang}%
 \email{aqiu@itp.ac.cn}

\affiliation{State Key Laboratory of Theoretical Physics,Institute of Theoretical Physics,\\
	Chinese Academy of Sciences,Beijing 100190, China.
	}

\begin{abstract}
The gravitational collapse of a massless scalar field with a self-interaction term $\lambda\phi^4$ in anti-de Sitter space is investigated.  We numerically investigate the effect of the self-interaction term on the critical amplitudes, forming time of apparent horizon, stable island and energy transformation. The results show that a positive $\lambda$ suppresses the formation of black hole, while a negative $\lambda$ enhances the process. We define two susceptibilities to characterize the effect of the self-interaction on the black hole formation, and find that near the critical amplitude, there exists a universal scaling relation with the critical exponent $\alpha \approx 0.74$ for the time of black hole formation.
\end{abstract}

\maketitle


\section{\label{sec:level1_1}Introduction}

Recently, a lot of attention has been focused to gravitational collapse in anti-de Sitter (AdS) spacetime. On the one hand, AdS spacetime is one of three maximal symmetric spacetimes (the other two are Minkowskian and de Sitter spacetimes). The gravitational
collapse in AdS spacetime is an interesting issue in its own right.  On the other hand,  AdS spacetime is a ground state of some superstring/M theories. Due to the so-called AdS/CFT correspondence, the form of black holes in AdS space is equivalent to the thermalization process of dual conformal field theory (CFT) in the AdS boundary.

However, due to the complexity of Einstein's field equations, it is quite hard to solve this problem analytically. Numerical methods and perturbation methods are usually employed for this goal. The pioneering numerical study on this issue finds that an arbitrary small spherically symmetric initial data of massless scalar field will collapse to form black hole~\cite{Bizon:2011gg,Maliborski:2013via,Buchel:2012uh}. Other class of perturbations leads to a similar conclusion~\cite{Dias:2011ss}. This implies that the AdS spacetime is unstable nonlinearly, quite different from the cases of Minkowskian and de Sitter spacetimes. Later on, exceptions were found in works~\cite{Buchel:2013uba,Maliborski:2013jca}, which show that there exist stable initial data immune to the black hole formation. Generalizations to massive scalar fields are investigated in~\cite{Kim:2014ida,Okawa:2015xma,Deppe:2015qsa}.

On the perturbation method side, improved perturbative expansions have been constructed to describe the small amplitude dynamics on time scales of order $1/\epsilon^2$, where $\epsilon$ denotes the amplitude of perturbation. The effective equations which describe the variations of the amplitudes and phases of AdS normal modes due to non-linearities are derived using multiscale~\cite{Balasubramanian:2014cja}, renormalization~\cite{Craps:2014vaa} and averaging~\cite{Craps:2014jwa} methods.

There are two crucial ingredients which are responsible for the instability of AdS space:
confinement property of AdS boundary and local non-linearity. To investigate the role played by local non-linearity, new non-linearity due to higher curvature terms has been studied in~\cite{Deppe:2014oua}. They found that the stability of AdS in 5D Einstein-Gauss-Bonnet gravity can be restored for small perturbations, due to the existence of mass gap of the Gauss-Bonnet black hole. Instead of adding new non-linear terms from gravity sector, in this paper, we consider the effect of self-interaction of matter field itself on the stability of AdS space. For simplicity, we consider the $\lambda\phi^4$ model of a massless scalar field.

The self-interaction has been considered in asymptotically flat spacetime~\cite{Okawa:2013jba}, which provides effective confinement and gives rise to some interesting phenomena. This is believed to have some connection with the massless scalar collapse in AdS spacetime. The energy flow between different modes on a fixed AdS background has been observed in~\cite{Basu:2014sia} with self-interaction providing the nonlinearity. Note that the energy flow is very important for the weakly turbulence instability of AdS spacetime. Therefore, the self-interaction provides another ``instability engine". These two engines may compete or cooperate with each other, which means that the self-interaction may enhance or suppress the instability of the system.

Our main motivation is to explore the effect of self-interaction on the instability of AdS spacetime. One of our main results is that a positive $\lambda$ suppresses the formation of black hole while a negative one promotes it. In addition, we find a universal scaling relation on the sensibility of  black hole formation time with respect to the self-interaction strength $\lambda$. The paper is organized as follows. In Section~\ref{sec:level1_2} we present the setup of the massless scalar collapse with self-interaction. Section~\ref{sec:level1_3} is devoted to our numerical results. We conclude in Section~\ref{sec:level1_4}.

\section{\label{sec:level1_2}Setup}

We consider a real massless scalar field with self-interaction in 3+1 dimensional spherically symmetric asymptotic AdS spacetime. The system is described by the following action,
\begin{equation}\label{action1}
    S=\int d^4x\sqrt{-g}\left[\frac1{16\pi G}(\mathcal{R}-2\Lambda)-\frac12(\nabla\phi)^2-\frac{\lambda}{4!}\phi^4\right],
\end{equation}
where $\mathcal{R}$ is the Ricci scalar curvature, $G$ is the Newtonian gravitation constant, and $\lambda$ is a real parameter, describing the self-interaction strength of the scalar field.  The Einstein's field  equation and the equation of motion of the scalar field read
\begin{align}
&\resizebox{.85\hsize}{!}{$ G_{\alpha\beta}+\Lambda g_{\alpha\beta}=8\pi G \left[\nabla_\alpha\phi \nabla_\beta\phi-\frac{1}{2} g_{\alpha\beta}(\nabla\phi)^2-\frac{\lambda}{4!}g_{\alpha\beta} \phi^4 \right]$},\label{eq:einstein} \\
& g^{\alpha\beta}\nabla_\alpha \nabla_\beta \phi-\frac{\lambda}6 \phi^3=0\, \label{eq:scalar} .
\end{align}

In order to find the solution of Eqs.~\eqref{eq:einstein}-\eqref{eq:scalar} in spherically symmetric AdS spacetime, we take the ansatz for the metric
\begin{align}
ds^2=\frac{\ell^2}{\cos^2 x}(-Ae^{-2\delta}dt^2+A^{-1}dx^2+\sin^2xd\Omega^2)\label{eq:Met Ansatz},
\end{align}
where $\ell^2=-3/\Lambda$ and $d\Omega^2$ is the standard metric on the round unit two-dimensional sphere, and $A, \delta$ and $\phi$ are all functions of time $t$ and radial coordinate $x$.

We now introduce two auxiliary variables $\Phi=\phi'$ and $\Pi=A^{-1}e^{\delta}\dot{\phi}$ and set $\ell=4\pi G=1$, then the equations of motion can be written as
\begin{align}
\dot \Phi =&\left( Ae^{-\delta}\Pi \right)',\label{eq:Phi}\\
\dot \Pi   =&\frac{1}{\tan^2x}\left( \tan^2xAe^{-\delta}\Phi \right)'-\frac{\lambda}{6}e^{-\delta} \sec^2x \phi^3, \label{eq:Pi}\\
A'        =&\frac{1+2\sin^2x}{\sin x \cos x}(1-A)\nonumber\\
             &-\sin x \cos x A\left(\Phi^2+\Pi^2\right)-\frac{\lambda}{12}\tan x\phi^4, \label{eq:A}\\
\delta' =&-\cos x \sin x \left(\Phi^2+\Pi^2\right), \label{eq:delta}\\
\phi'=&\Phi, \label{eq:phi}
\end{align}
where an overdot stands for the derivative with respect to time $t$ and a prime to the radial coordinate $x$.
This model shares the same boundary conditions as the case without the self-interaction~\cite{Bizon:2011gg}. At the origin $x=0$,
\begin{equation}\label{bc:center}
\begin{split}
\phi(t,x)=&f_0(t)+\mathcal{O}(x^2),\\
\delta(t,x)=&\mathcal{O}(x^2),\\
A(t,x)=&1+\mathcal{O}(x^2).
\end{split}
\end{equation}
and at the spatial infinity $x=\pi/2$,
\begin{equation}\label{bc:boundary}
\begin{split}
\phi(t,x)=&f_\infty(t)\rho^3+\mathcal{O}(\rho^5),\\
\delta(t,x)=&\delta_\infty(t)+\mathcal{O}(\rho^6),\\
A(t,x)=&1-2M\rho^3+\mathcal{O}(\rho^6),
\end{split}
\end{equation}
where $\rho=\pi/2-x$.

The $\lambda\phi^4$ model appears in many toy models to study the effects of the self-interaction. For quantum field theory in flat spacetime, we need $\lambda\geq0$ so that the scalar field has a stable ground state. In asymptotic AdS space-time, because the asymptotic boundary at $x=\pi/2$ gives an infinite high potential barrier, the massless scalar field can still be stable when $\lambda<0$. In this paper, we will consider both $\lambda>0$ and $\lambda<0$ to see how these two kinds of interactions influence the evolution of the scalar field in AdS space.

\section{\label{sec:level1_3}Numerical Results}

Following~\cite{Maliborski:2013via}, we use a 4th-order Runge-Kutta method to solve the time evolution equations \eqref{eq:Phi}-\eqref{eq:Pi}. At each time step, the metric functions $A$ and $\delta$ are given by integrating the constraint equations \eqref{eq:A}-\eqref{eq:delta} from the origin to the infinity also using a 4th-order Runge-Kutta method, while the scalar field $\phi$ is given by integrating the equation~\eqref{eq:phi} backward from the infinity to the origin. In order to clearly
 see the influence of the self-interaction, we also take the Gaussian initial data for the  scalar field as
\begin{align}
\Phi(0,x)=0,\ \ \Pi(0,x)=\epsilon \exp\left(-\frac{\tan^2x}{\sigma^2}\right), \label{init}
\end{align}
where $\epsilon$ stands for the amplitude, while $\sigma$ for the width of the initial wave packet.

Under this class of initial data, the stability of space-time depends on whether a black hole could form after some time, which is
signalled by the appearance of an apparent horizon at a point $x_H$ where $A(t, x)$ falls into zero. There exists strong evidence that the initial data are classified into two categories: unstable states and stable states. For those unstable states, the wave configuration oscillates between the origin and the boundary a few times, then collapses to black hole. For those stable states, the wave configuration stays regular everywhere in the cavity all the time. This kind of states is often referred to as ``stable island" in the initial data phase space. There are three parameters in this system: self-interaction strength $\lambda$, amplitude $\epsilon$ and width $\sigma$ of the initial data. We will study  the effect of self-interaction on these two kinds of states separately for different amplitude $\epsilon$ and width $\sigma$.

\subsection{\label{sec:level2_1}Effect on unstable states}

To see the effect of the self-interacting term on the instability of the system, we first fix width $\sigma=1/16$, and try to find out the influence of $\lambda$ on the unstable states with different $\epsilon$. Then we try to find out the influence of self-interaction on the unstable states with a few different $\sigma$.

\subsubsection{\label{sec:level3_1} The case with $\sigma=1/16$}

Since the apparent horizon is formed very close to the center of the space, the time difference of black hole formation due to the different strength of the self-interaction is diluted by the travel time over the whole cavity (from the origin to the boundary). So the time of black hole formation is still dominated by the time of the scalar field oscillation in the cavity. As a result, it is expected that the time difference caused by the self-interaction is small in general.

\begin{figure}
	\includegraphics[width=90mm,clip]{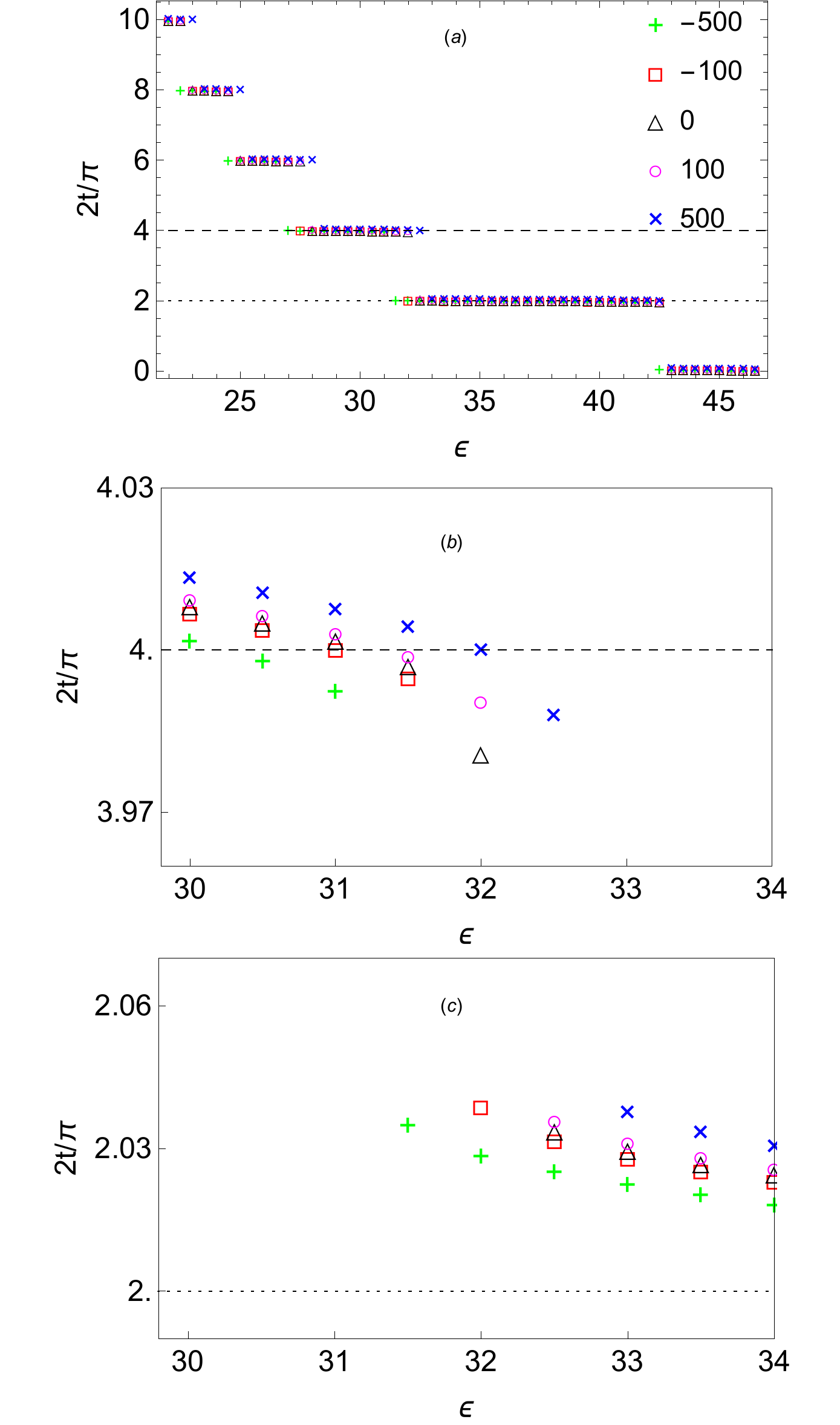}
	\caption{Formation time of black hole from scalar field with different self-interaction strength $\lambda$, we set $\sigma=1/16$. In the top panel, we plot the formation time $t$ with respect to amplitude $\epsilon$ for different $\lambda$. The middle and bottom panels show the formation time $t$ when $\epsilon\in[30,34]$ for different $\lambda$, respectively.}
	\label{fig:formation_time}
\end{figure}

We fix the width of the initial data $\sigma=1/16$ and set $\lambda=-500,-100,0,100,500$, respectively. We show the influence of self-interaction on the formation time of black hole in Fig.~\ref{fig:formation_time}. The general behavior of the black hole formation time is similar to the case of $\lambda=0$ for every $\lambda$. As the amplitude $\epsilon$ decreases, it approximately forms ascending steps and increases monotonically on every step. On every step, the black hole formation time is almost the same for every $\lambda$. But as we improve the resolution, which is shown in the middle and bottom plots in Fig.~\ref{fig:formation_time}, we can see the difference. Fig.~\ref{fig:formation_time}(b) shows the case around $\epsilon\in[30, 34]$ and $t\simeq4\times \frac{\pi}{2}$, while Fig.~\ref{fig:formation_time}(c) shows the case in the same $\epsilon$ region but $t\simeq2\times\frac{\pi}{2}$ (which are around the second critical amplitude $\epsilon_1\simeq32.5$). If we fix the amplitude, the black hole formation time is decreased with the decrease of $\lambda$, although very small. However, around the critical amplitude, this time difference between different $\lambda$ can be huge. We look at the critical amplitude around $\epsilon\sim 31.5$, for instance. As we can see in Fig~.\ref{fig:formation_time}(b), when $\epsilon\lesssim 31$, the formation time decreases a little as we decrease $\lambda$. In this case, the time difference between $\lambda=-500$ and $\lambda=0$ is roughly $0.006\times\frac{\pi}{2}$. When $\epsilon\gtrsim 31.5$, some of the initial data disappear in Fig.~\ref{fig:formation_time}(b) and jump to the previous step shown in Fig.~\ref{fig:formation_time}(c). The smaller the self-interaction coefficient, the faster the jump happens. These jumps cause huge time difference between different $\lambda$. In this case, the time difference between $\lambda=-500$ and $\lambda=0$ could be more than $2\times\frac{\pi}{2}$. When $\epsilon$ is large enough ($\epsilon > 32.5$ in this case), all the formation times jump to the previous step. The similar time jump of black hole formation also happens in the case of $\lambda>0$.

This kind of time jump caused by $\lambda$ near the critical amplitude can be understood as follows. When $\lambda <0$, the self-interaction enhances the instability of the system, which makes the black hole formation a bit earlier than the case without the self-interaction. When this small time shift happens near the critical point, it may push the black hole formation out of the effective concentration region which is a small region very close to the origin of the space. This means that the black hole formation has to occur in the previous effective concentration region, which causes a huge time jump (earlier). When $\lambda >0$, the situation is just opposite. The self-interaction makes the formation of black hole a bit later. When this suppressing effect happens near the critical amplitude, it may pull the formation of black hole out of the effective concentration region and make the scalar oscillate one more time in the cavity. It causes a huge time delay in the formation of black hole.

Besides the time for the appearance of apparent horizon is influenced by the self-interaction, the critical amplitudes $\epsilon_n$ which give the zero apparent horizon radius are also shifted by $\lambda$, though this shift is very small in the first few critical amplitudes. From Figs. (b) and (c) in Fig.~\ref{fig:formation_time}, we see that the second critical amplitude $\epsilon_1$ is increased for positive $\lambda$ but decreased for negative $\lambda$. For a given integer $n\geq0$, the critical amplitude $\epsilon_n$ is an increasing function of $\lambda$ in the region our numerical computation can cover. This critical amplitude shift with $\lambda$ is consistent with the behavior of the time shift of black hole formation in the previous paragraph. To explore this, one can suppose that, for a given $\lambda=\lambda_0$, there is amplitude $\epsilon'$ which is larger but very close to a critical amplitude $\epsilon_i$ and gives the apparent horizon radius $x_H(\epsilon', \lambda_0)$ is very close to zero. Now suppose we alter $\lambda$ a little such as $\lambda=\lambda+\delta\lambda$ with $\delta\lambda>0$. Because the larger $\lambda$ will lead the apparent horizon to appear later, the peak of the $\Phi$ and $\Pi$ or the bottom of $A$ can propagate into the region closer to the origin  before an apparent horizon appears, which leads the apparent horizon radius to be smaller. By adjusting the value of $\delta\lambda$, we can make apparent horizon radius decrease to zero and $\epsilon'$ is a new critical amplitude. We see that by increasing the value of $\lambda$, the new critical amplitude is larger than the old one.

\subsubsection{\label{sec:level3_2} The case with different $\sigma$}

\begin{figure}
	\includegraphics[width=90mm,clip]{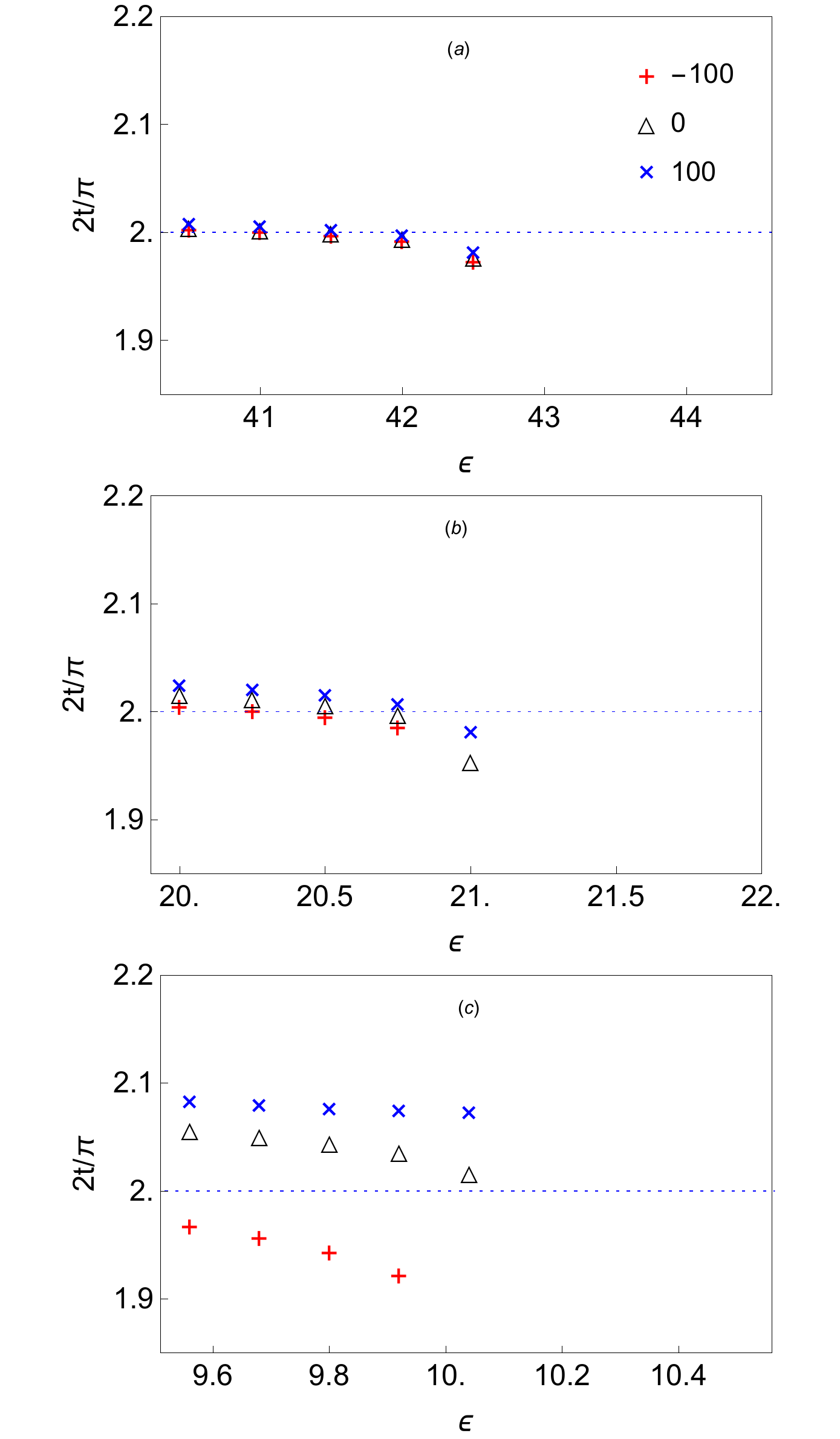}
	\caption{Formation time of black hole from scalar field collapse with different initial widthes. Top: $\sigma=1/16$, Middle: $\sigma=1/8$, Bottom: $\sigma=1/4$. We set $\lambda=-100,0,100$}
	\label{fig:formation_time_sig}
\end{figure}

In this subsection, we fix $\lambda=-100,0,100$, and consider the width of initial data as $\sigma=1/16,1/8,1/4$, respectively. In order to see the influence of self-interaction on the gravitational collapse, we magnify the region around the first critical amplitude ($\epsilon_0$), while Fig.~\ref{fig:formation_time} magnifies the region around the second critical amplitude ($\epsilon_1$). The results are shown in Fig.~\ref{fig:formation_time_sig}.

Qualitatively, the influence of self-interaction with different initial widths is the same. It enhances (when $\lambda<0$) or suppresses (when $\lambda>0$) the formation of black hole. Quantitatively, there exist differences. Figs.~\ref{fig:formation_time_sig}(a-c) show the black hole formation times for the initial data with widthes $\sigma=1/16,1/8,1/4$, respectively. When $\sigma=1/16$, the time difference between the two cases with a nonzero $\lambda$ and vanishing $\lambda$ is very small, less than $0.005\times\frac{\pi}{2}$. As we increase the width of initial data, the enhancement (or suppressing) effect caused by the same strength of self-interaction becomes obvious. When $\sigma=1/8$, the time difference between the two cases with a nonzero $\lambda$ and vanishing $\lambda$ is bigger, and reaches about $0.01\times\frac{\pi}{2}$. When $\sigma=1/4$, it is more obvious, the time difference is around $0.05\times\frac{\pi}{2}$. The time difference is much obvious for the case with a negative $\lambda$. 
%


The above observation can be understood as follows. When $\sigma$ is small, the wave packet decays rapidly in space, which means that the self-interaction only happens in a very narrow region and gives very weak influence on the system. In this case, one can expect that the time differences of black hole formation for different $\lambda$ are very small. However, when we increase the value of $\sigma$, the region where the self-interaction plays its role is enlarged, so its influence on the system become stronger.  In this case, one can expect that the time differences for different $\lambda$ become obvious.

\subsubsection{\label{sec:level3_3}Sensibilities}

To characterize the influence of the self-interaction on the critical amplitude and time of black hole formation with respect to the self-interaction strength $\lambda$ under the initial data \eqref{init}, we define two sensibility coefficients,
\begin{equation}\label{sensib}
    \chi_{\epsilon_n}(\sigma)=\lim_{\lambda\rightarrow0}\left(\frac{\partial\epsilon_n}{\partial\lambda}\right)_{\sigma}, ~~\chi_{t}(\epsilon,\sigma)=\lim_{\lambda\rightarrow0}\left(\frac{\partial t}{\partial\lambda}\right)_{\sigma,\epsilon}.
\end{equation}
The first one in Eqs.~\eqref{sensib} describes the sensibility of critical amplitude and the second one  describes the sensibility of forming time of black hole.

In Table.~\ref{tab:dadl}, we list the first four $\chi_{\epsilon_n}(\sigma)$ when $\sigma=1/16$. One can see that the values of $\chi_{\epsilon_n}$ are all positive, which is in agreement with our numerical calculations that a larger $\lambda$ leads to a larger critical amplitude.  In addition, we see that the values of $\chi_{\epsilon_n}$ decrease with $n$ (the exception is the case with $n=0$). This shows the fact that the sensibility of critical amplitude of scalar field to $\lambda$ is decreased with $n$.

\begin{table}
\centering
\begin{tabular}{*{5}{|c}|}\hline
$\epsilon_n$        & $\epsilon_0$ & $\epsilon_1$ & $\epsilon_2$ & $\epsilon_3$ \\\hline
$\chi_{\epsilon_n}\times 10^{3}$ & 0.15    & 1.2      & 1.0     & 0.90 \\\hline
\end{tabular}
\caption{Sensibility of critical amplitude $\epsilon_n$ to $\lambda$ when $\sigma=1/16$}
\label{tab:dadl}
\end{table}

\begin{figure}
	\includegraphics[width=80mm,clip]{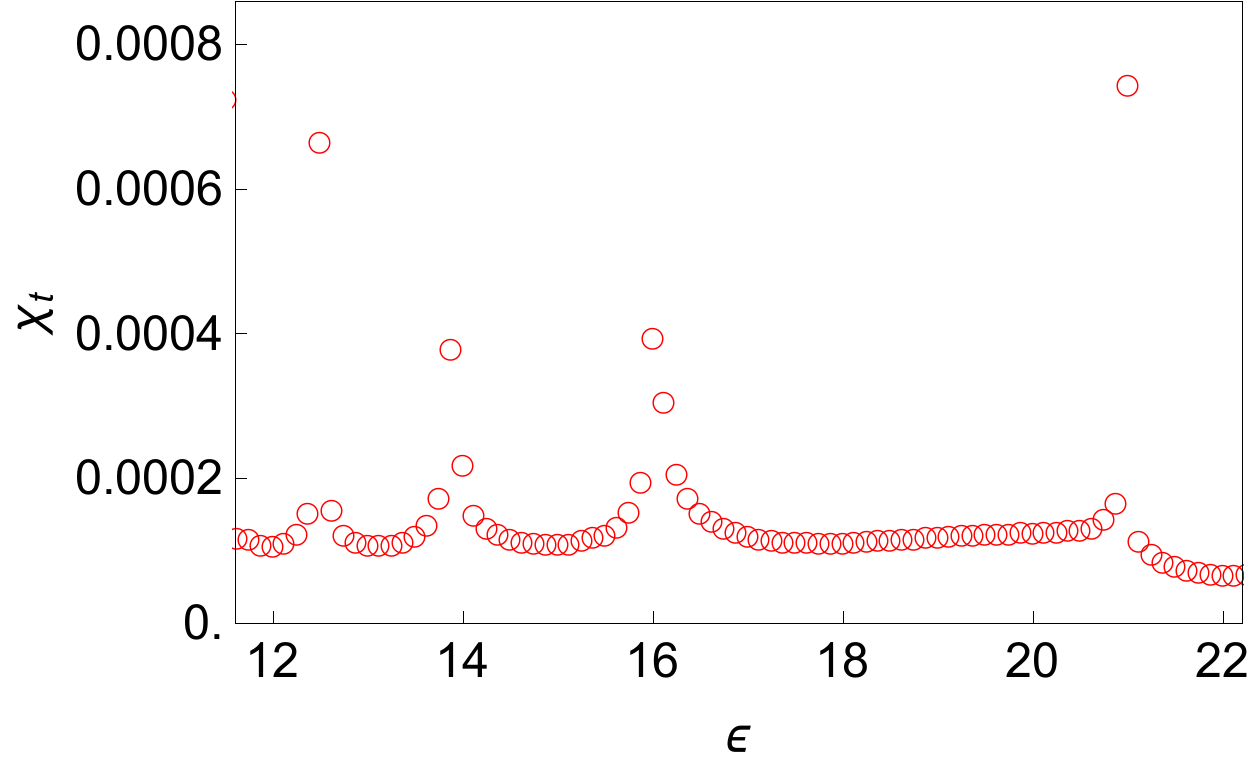}
\includegraphics[width=80mm,clip]{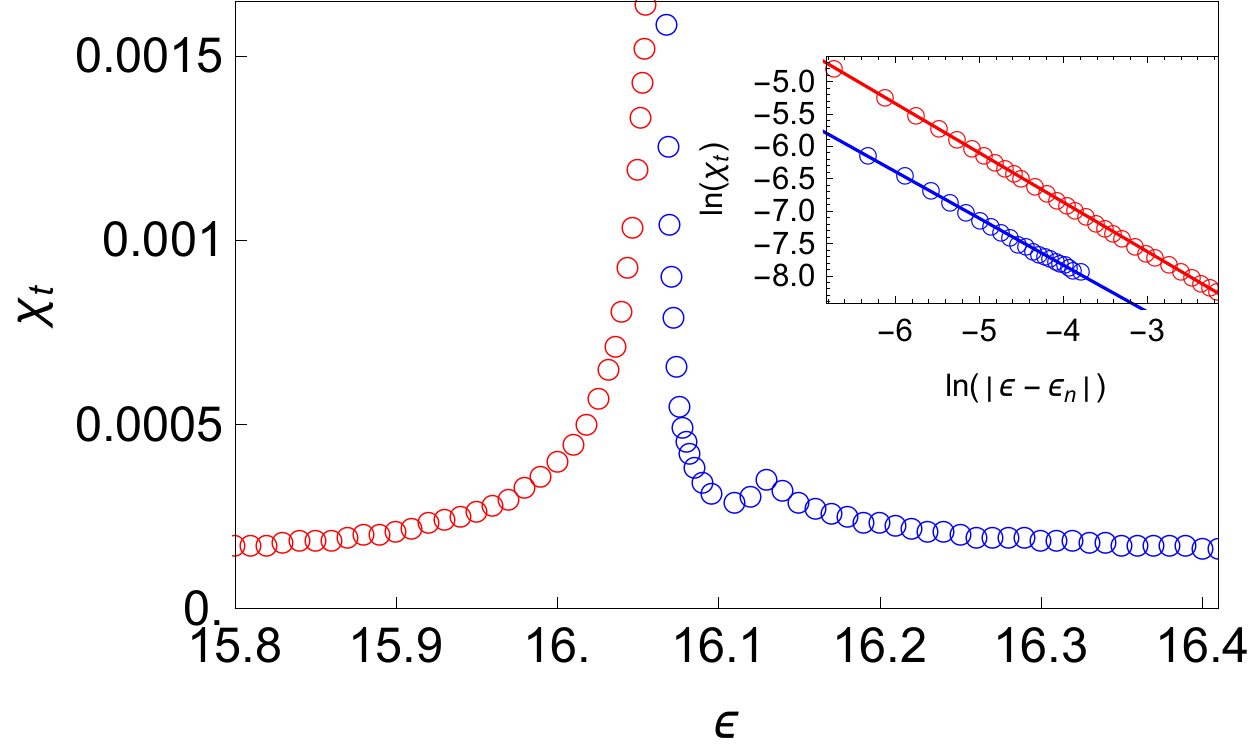}
	\caption{The relation of $\chi_{t}(\epsilon,\sigma)$ with respect to $\epsilon$ when $\sigma=1/8$. In the upper plot, we scan $\epsilon$ from 11.25 to 22, and show that there is a pole for every $\epsilon=\epsilon_n$. In the bottom plot, we show the value of $\chi_{t}(\epsilon,\sigma)$ around $\epsilon=\epsilon_1$. The inset in the bottom  plot shows the fitting curves using Eqs.~\eqref{scaling1}. The blue is the case $\epsilon\rightarrow\epsilon_1^+$, while the red is the case $\epsilon\rightarrow\epsilon_1^-$.}
	\label{fig:scaling}
\end{figure}

One of very interesting results by including the $\lambda\phi^4$ from our numerical computations is about $\chi_{t}(\epsilon,\sigma)$. By the definition in \eqref{sensib}, we can see that $\chi_{t}(\epsilon)$ diverges when $\epsilon=\epsilon_n$. Near the critical amplitude $\epsilon\rightarrow\epsilon_n$, we observe a scaling relation as,
\begin{equation}\label{scaling1}
    \chi_{t}(\epsilon,\sigma)\propto\left\{
    \begin{split}
    \frac1{(\epsilon_n-\epsilon)^\alpha},~~&\epsilon\rightarrow\epsilon_n^-\\
    \frac1{(\epsilon-\epsilon_n)^{\alpha'}},~~&\epsilon\rightarrow\epsilon_n^+.
    \end{split}
    \right.
\end{equation}
 Fig.~\eqref{fig:scaling} shows the relation of $\chi_{t}(\epsilon,\sigma)$ with respect to $\epsilon$ when $\sigma=1/8$. In the upper plot, we scan  $\epsilon$ from 11.25 to 22. As is expected, $\chi_{t}(\epsilon,\sigma)$ is always positive and a pole appears for every $\epsilon=\epsilon_n$. In the bottom plot, we show the value of $\chi_{t}(\epsilon,\sigma)$ around $\epsilon=\epsilon_1$. By this figure, we can see it clear that when $\epsilon$ is near to its critical value, the system is very sensitive to the self-interaction $\lambda\phi^4$ term. At the critical amplitude, an infinitesimal $\lambda \phi^4$ term can give rise to a very essential difference. This is not very surprising. Because there is a naked singularity at the center of the space when $\epsilon=\epsilon_n$, which will lead to the breaking of causality and stability of the spacetime~\cite{Christodoulou}.  To find the values of $\alpha$ and $\alpha'$ in Eqs.~\eqref{scaling1}, we fit the values of $\chi_{t}(\epsilon,\sigma)$ when $\epsilon\rightarrow\epsilon_n$ for different $n$. We find that $\alpha\simeq\alpha'\simeq0.74(2)$, which are independent on $n$ and $\sigma$, up to numerical errors.

\subsection{\label{sec:level2_2}Effect on stable states}

``Stable islands" have been claimed to exist in the free scalar case. We want to see whether the self-interaction plays  any role on these ``islands" in the initial data. For simplicity, we here consider only the ``island" with a large $\sigma$~\cite{Buchel:2013uba}.

\begin{figure}
	\includegraphics[width=90mm,clip]{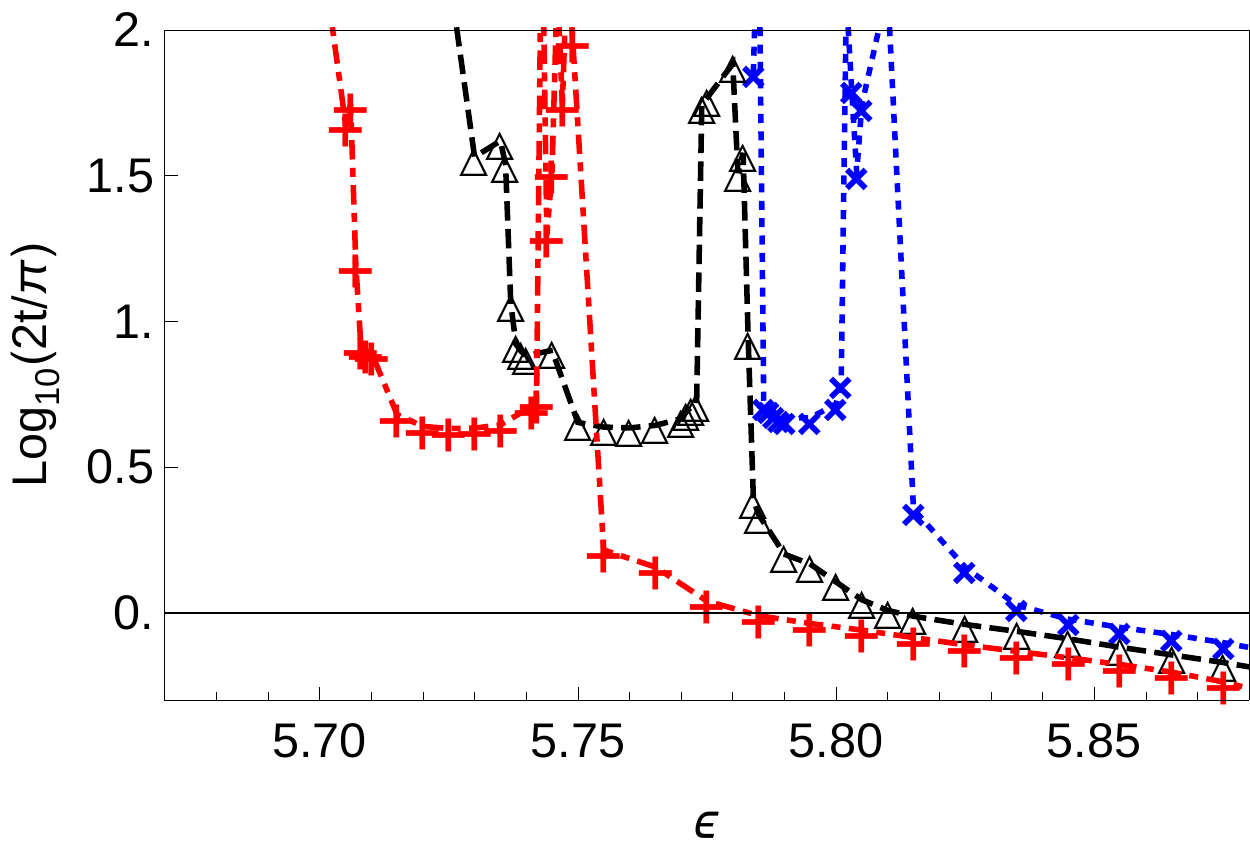}
	\caption{Formation time of black hole from scalar filed collapse with large width $\sigma=2/5$.  Red: $\lambda=-10$, Black: $\lambda=0$, Blue: $\lambda=10$}
	\label{fig:large_sig}
\end{figure}

We set the width of initial data $\sigma=2/5$ . The result is shown in Fig.~\ref{fig:large_sig} for three different self-interaction strengthes: $\lambda=-10, 0, 10$, respectively.
We can see that all these three initial data sets show similar behavior for the black hole formation time and that there is a transition from the black hole formation phase to stable phase as $\epsilon$ decreases. We also notice the existence of the ``bump", as in~\cite{Buchel:2013uba}, in the black hole formation time before it grows monotonically with the decrease of $\epsilon$.

When the self-interaction of the scalar field is not vanishing, the bump is shifted. When $\lambda=10$, the center of the bump is around $\epsilon\simeq5.81$ which is larger than the case of $\lambda=0$ whose bump is centered around $\epsilon\simeq 5.78$. We believe that the shift of the bump is a sign of expansion of the size of ``stable island" due to positive $\lambda$. When $\lambda=-10$, the situation is opposite. The center of the bump is around $5.75$, which indicates the size of ``stable island" shrinks due to a negative $\lambda$.

The expansion or shrink of the size of ``stable island" is an indication of suppressing or enhancing the instability of the system. So the self-interaction has the same effect on stable states as on unstable states: positive $\lambda$ suppresses the instability of the system, while negative $\lambda$ enhances it.

\subsection{Effect on the energy transfer}{\label{sec:level2_3}

For small amplitude $\epsilon$, the system can evolute a very long time before a trapped surface forms which indicts the appearance of apparent horizon. To see further the influence of $\lambda\phi^4$ on the instability of AdS space, in this section, following Ref.~\cite{Bizon:2011gg}, we investigate the energy transfer between different modes.

In the case with small amplitude $\epsilon$, we can expand the functions $\{\phi, A, \delta\}$ as,
\begin{equation}\label{expand1}
    \phi=\sum_{j=0}^{\infty}\phi_{2j+1}\epsilon^{2j+1},~A=1-\sum_{j=1}^{\infty}A_{2j}\epsilon^{2j},~\delta=\sum_{j=1}^{\infty}\delta_{2j}\epsilon^{2j}.
\end{equation}
Then at the linear order of $\epsilon$, the solutions of Eqs.~\eqref{eq:Phi}-\eqref{eq:phi} are $A=1, \delta=0$ and $\phi$ can be expressed by hypergeometric function such as~\cite{Bizon:2011gg},
\begin{equation}\label{expphi2}
    \phi_j=a_j\cos(\omega_jt+\beta_j)e_j(x)
\end{equation}
with some constants $a_j, \beta_j$ and
\begin{equation}\label{hyperge1}
    e_j(x)=d_j\cos^3x {~_2F_1}(-j,3+j,\frac32;\sin^2x).
\end{equation}
Here $\omega_j=\pm(3+2j)$ and $d_j=\sqrt{16(j+1)(j+2)/\pi}$ with $j=0,1,2,\cdots$.

Using the linear order solutions~\eqref{expphi2}, we can project a general solution $\{\Phi, \Pi\}$(not only in the linear order of $\epsilon$) as,
\begin{equation}\label{project1}
    \Phi_j= \langle\sqrt{A}\Phi,e_j'\rangle,~\Pi_j=\langle\sqrt{A}\Pi,e_j\rangle
\end{equation}
Here the inner product is defined as $\langle f,g\rangle=\int_0^{\pi/2}f(x)g(x)\tan^2xdx$. Then the energy of $j$-mode can be expressed as,
\begin{equation}\label{energyj}
    E_j=\Pi_j^2+\omega^{-2}_j\Phi_j^2.
\end{equation}

To investigate the influence of $\lambda\phi^4$ term on the energy transfer, we use the two modes initial data as in Ref.~\cite{Bizon:2011gg}, i.e., $\phi(0,x)=\epsilon[e_0(x)/d_0+e_1(x)/d_1]$ and define,
\begin{equation}\label{Dsimga1}
    \Delta_k(\lambda_0)=(\sum_{i=0}^kE_i)|_{\lambda=\lambda_0}-(\sum_{i=0}^kE_i)|_{\lambda=0}.
\end{equation}
For a given $\lambda_0$, $\Delta_k(\lambda_0)$ describes the difference of the energy staying in the first $k$ models between the cases with $\lambda_0 \ne 0$ and with $\lambda=0$. Thus if it is negative, it means that the $\lambda\phi^4$ term can accelerate the energy transfer into high energy modes, and vice versa.
\begin{figure}
	\includegraphics[width=90mm,clip]{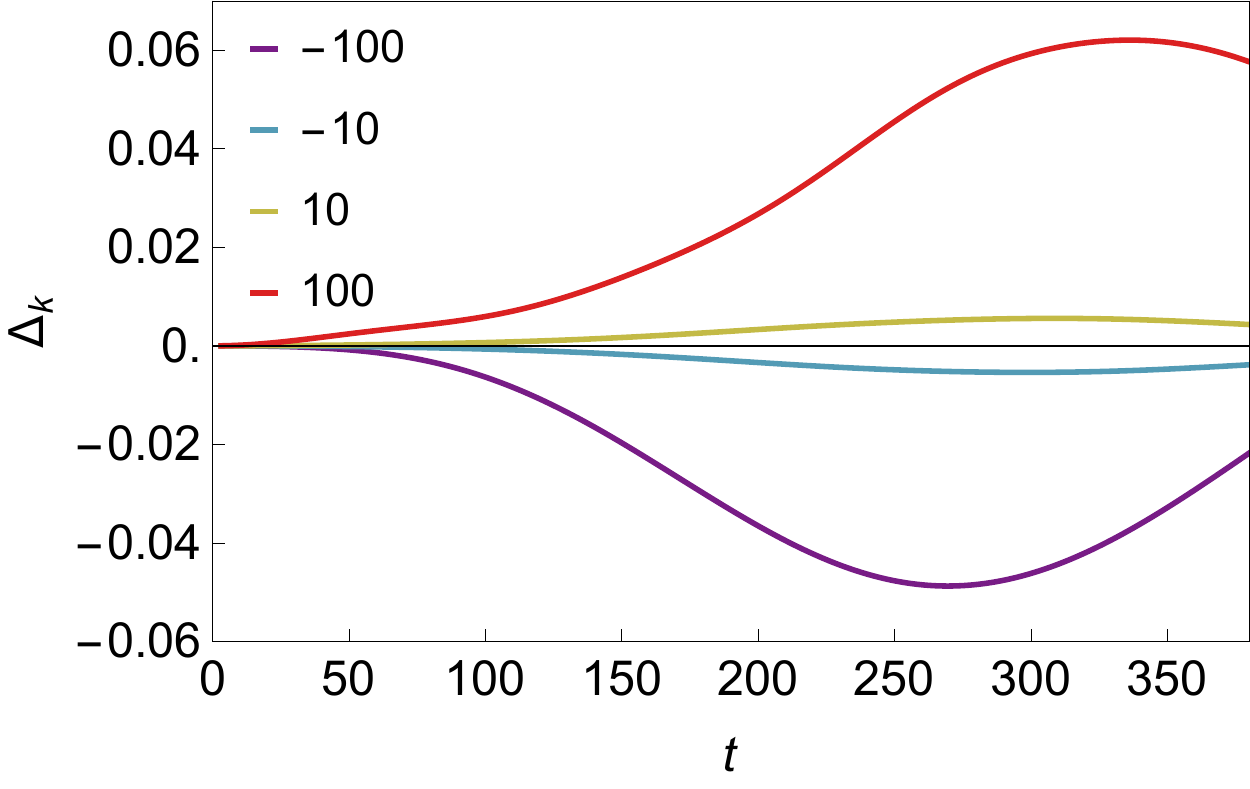}
	\caption{The value of $\Delta_k$ for different $\lambda$ with large width $\sigma=2/5$. Here we take $k=2$.}
	\label{energyt}
\end{figure}

In Fig.~\ref{energyt}, we plot $\Delta_k(\lambda_0)$ when $\epsilon=0.088$ for different $\lambda_0$ and show the result for $\lambda=-100,-10,10$ and $100$, respectively. It can be clearly seen that $\Delta(\lambda_0)$ is positive when $\lambda>0$, which means that a positive $\lambda$ can enhance the stability and make the energy stay in low energy modes much long. We can expect the in this case it will lead the black hole to form later than the case of $\lambda=0$, which is consistent with our numerical computation in the case with large amplitudes.

\section{\label{sec:level1_4}Conclusion}

We have studied the gravitational collapse of massless scalar field with a self-interaction $\lambda \phi^4$ in AdS space, paying attention on the
influence of the self-interaction on the instability of AdS space. This self-interaction leads to an enhancing ($\lambda<0$) or suppressing ($\lambda>0$) effect on the formation of black hole. We have seen that near the critical amplitude $\epsilon$, this self-interaction may cause a large time difference of black hole formation between free scalar field case and self-interacting scalar field case (oscillating one more or one less in the cavity). We have defined two susceptibilities to characterize the effect of the self-interaction, one is the amplitude with respect to the self-interaction strength $\lambda$, the other is the formation time of
black hole. We have found a universal scaling relation for the formation time of black hole near the critical amplitude, which is found independent of $n$ and $\sigma$; the critical exponent $\alpha \approx 0.74$. We have also investigated the effect of $\lambda\phi^4$ on the energy transfer. The results show that a positive $\lambda$ will delay energy transfer into high energy modes, while a negative $\lambda$ can accelerate this transfer.  In addition, we have studied the effect of the self-interaction on the ``stable island" in the initial data with a large $\sigma$, and found that a positive (negative) $\lambda$ expands (shrinks) the size of the `` stable island" and leads to a shift of the critical amplitude.

\begin{acknowledgements}
 This work was supported in part by the National Natural Science Foundation of China
(No.11375247 and No.11435006 ).

\end{acknowledgements}



\vskip 5cm

\bibliography{self_int_refs}

\begin{thebibliography}{16}%
\makeatletter
\providecommand \@ifxundefined [1]{%
 \@ifx{#1\undefined}
}%
\providecommand \@ifnum [1]{%
 \ifnum #1\expandafter \@firstoftwo
 \else \expandafter \@secondoftwo
 \fi
}%
\providecommand \@ifx [1]{%
 \ifx #1\expandafter \@firstoftwo
 \else \expandafter \@secondoftwo
 \fi
}%
\providecommand \natexlab [1]{#1}%
\providecommand \enquote  [1]{``#1''}%
\providecommand \bibnamefont  [1]{#1}%
\providecommand \bibfnamefont [1]{#1}%
\providecommand \citenamefont [1]{#1}%
\providecommand \href@noop [0]{\@secondoftwo}%
\providecommand \href [0]{\begingroup \@sanitize@url \@href}%
\providecommand \@href[1]{\@@startlink{#1}\@@href}%
\providecommand \@@href[1]{\endgroup#1\@@endlink}%
\providecommand \@sanitize@url [0]{\catcode `\\12\catcode `\$12\catcode
  `\&12\catcode `\#12\catcode `\^12\catcode `\_12\catcode `\%12\relax}%
\providecommand \@@startlink[1]{}%
\providecommand \@@endlink[0]{}%
\providecommand \url  [0]{\begingroup\@sanitize@url \@url }%
\providecommand \@url [1]{\endgroup\@href {#1}{\urlprefix }}%
\providecommand \urlprefix  [0]{URL }%
\providecommand \Eprint [0]{\href }%
\providecommand \doibase [0]{http://dx.doi.org/}%
\providecommand \selectlanguage [0]{\@gobble}%
\providecommand \bibinfo  [0]{\@secondoftwo}%
\providecommand \bibfield  [0]{\@secondoftwo}%
\providecommand \translation [1]{[#1]}%
\providecommand \BibitemOpen [0]{}%
\providecommand \bibitemStop [0]{}%
\providecommand \bibitemNoStop [0]{.\EOS\space}%
\providecommand \EOS [0]{\spacefactor3000\relax}%
\providecommand \BibitemShut  [1]{\csname bibitem#1\endcsname}%
\let\auto@bib@innerbib\@empty
\bibitem [{\citenamefont {Bizon}\ and\ \citenamefont
  {Rostworowski}(2011)}]{Bizon:2011gg}%
  \BibitemOpen
  \bibfield  {author} {\bibinfo {author} {\bibfnamefont {P.}~\bibnamefont
  {Bizon}}\ and\ \bibinfo {author} {\bibfnamefont {A.}~\bibnamefont
  {Rostworowski}},\ }\href {\doibase 10.1103/PhysRevLett.107.031102} {\bibfield
   {journal} {\bibinfo  {journal} {Phys. Rev. Lett.}\ }\textbf {\bibinfo
  {volume} {107}},\ \bibinfo {pages} {031102} (\bibinfo {year} {2011})},\
  \Eprint {http://arxiv.org/abs/1104.3702} {arXiv:1104.3702 [gr-qc]}
  \BibitemShut {NoStop}%
\bibitem [{\citenamefont {Maliborski}\ and\ \citenamefont
  {Rostworowski}(2013{\natexlab{a}})}]{Maliborski:2013via}%
  \BibitemOpen
  \bibfield  {author} {\bibinfo {author} {\bibfnamefont {M.}~\bibnamefont
  {Maliborski}}\ and\ \bibinfo {author} {\bibfnamefont {A.}~\bibnamefont
  {Rostworowski}},\ }\bibfield  {booktitle} {\emph {\bibinfo {booktitle}
  {{Proceedings, Spring School on Numerical Relativity and High Energy Physics
  (NR/HEP2)}}},\ }\href {\doibase 10.1142/S0217751X13400204} {\bibfield
  {journal} {\bibinfo  {journal} {Int. J. Mod. Phys.}\ }\textbf {\bibinfo
  {volume} {A28}},\ \bibinfo {pages} {1340020} (\bibinfo {year}
  {2013}{\natexlab{a}})},\ \Eprint {http://arxiv.org/abs/1308.1235}
  {arXiv:1308.1235 [gr-qc]} \BibitemShut {NoStop}%
\bibitem [{\citenamefont {Buchel}\ \emph {et~al.}(2012)\citenamefont {Buchel},
  \citenamefont {Lehner},\ and\ \citenamefont {Liebling}}]{Buchel:2012uh}%
  \BibitemOpen
  \bibfield  {author} {\bibinfo {author} {\bibfnamefont {A.}~\bibnamefont
  {Buchel}}, \bibinfo {author} {\bibfnamefont {L.}~\bibnamefont {Lehner}}, \
  and\ \bibinfo {author} {\bibfnamefont {S.~L.}\ \bibnamefont {Liebling}},\
  }\href {\doibase 10.1103/PhysRevD.86.123011} {\bibfield  {journal} {\bibinfo
  {journal} {Phys. Rev.}\ }\textbf {\bibinfo {volume} {D86}},\ \bibinfo {pages}
  {123011} (\bibinfo {year} {2012})},\ \Eprint {http://arxiv.org/abs/1210.0890}
  {arXiv:1210.0890 [gr-qc]} \BibitemShut {NoStop}%
\bibitem [{\citenamefont {Dias}\ \emph {et~al.}(2012)\citenamefont {Dias},
  \citenamefont {Horowitz},\ and\ \citenamefont {Santos}}]{Dias:2011ss}%
  \BibitemOpen
  \bibfield  {author} {\bibinfo {author} {\bibfnamefont {O.~J.~C.}\
  \bibnamefont {Dias}}, \bibinfo {author} {\bibfnamefont {G.~T.}\ \bibnamefont
  {Horowitz}}, \ and\ \bibinfo {author} {\bibfnamefont {J.~E.}\ \bibnamefont
  {Santos}},\ }\href {\doibase 10.1088/0264-9381/29/19/194002} {\bibfield
  {journal} {\bibinfo  {journal} {Class. Quant. Grav.}\ }\textbf {\bibinfo
  {volume} {29}},\ \bibinfo {pages} {194002} (\bibinfo {year} {2012})},\
  \Eprint {http://arxiv.org/abs/1109.1825} {arXiv:1109.1825 [hep-th]}
  \BibitemShut {NoStop}%
\bibitem [{\citenamefont {Buchel}\ \emph {et~al.}(2013)\citenamefont {Buchel},
  \citenamefont {Liebling},\ and\ \citenamefont {Lehner}}]{Buchel:2013uba}%
  \BibitemOpen
  \bibfield  {author} {\bibinfo {author} {\bibfnamefont {A.}~\bibnamefont
  {Buchel}}, \bibinfo {author} {\bibfnamefont {S.~L.}\ \bibnamefont
  {Liebling}}, \ and\ \bibinfo {author} {\bibfnamefont {L.}~\bibnamefont
  {Lehner}},\ }\href {\doibase 10.1103/PhysRevD.87.123006} {\bibfield
  {journal} {\bibinfo  {journal} {Phys. Rev.}\ }\textbf {\bibinfo {volume}
  {D87}},\ \bibinfo {pages} {123006} (\bibinfo {year} {2013})},\ \Eprint
  {http://arxiv.org/abs/1304.4166} {arXiv:1304.4166 [gr-qc]} \BibitemShut
  {NoStop}%
\bibitem [{\citenamefont {Maliborski}\ and\ \citenamefont
  {Rostworowski}(2013{\natexlab{b}})}]{Maliborski:2013jca}%
  \BibitemOpen
  \bibfield  {author} {\bibinfo {author} {\bibfnamefont {M.}~\bibnamefont
  {Maliborski}}\ and\ \bibinfo {author} {\bibfnamefont {A.}~\bibnamefont
  {Rostworowski}},\ }\href {\doibase 10.1103/PhysRevLett.111.051102} {\bibfield
   {journal} {\bibinfo  {journal} {Phys. Rev. Lett.}\ }\textbf {\bibinfo
  {volume} {111}},\ \bibinfo {pages} {051102} (\bibinfo {year}
  {2013}{\natexlab{b}})},\ \Eprint {http://arxiv.org/abs/1303.3186}
  {arXiv:1303.3186 [gr-qc]} \BibitemShut {NoStop}%
\bibitem [{\citenamefont {Kim}(2015)}]{Kim:2014ida}%
  \BibitemOpen
  \bibfield  {author} {\bibinfo {author} {\bibfnamefont {N.}~\bibnamefont
  {Kim}},\ }\href {\doibase 10.1016/j.physletb.2015.01.045} {\bibfield
  {journal} {\bibinfo  {journal} {Phys. Lett.}\ }\textbf {\bibinfo {volume}
  {B742}},\ \bibinfo {pages} {274} (\bibinfo {year} {2015})},\ \Eprint
  {http://arxiv.org/abs/1411.1633} {arXiv:1411.1633 [hep-th]} \BibitemShut
  {NoStop}%
\bibitem [{\citenamefont {Okawa}\ \emph {et~al.}(2015)\citenamefont {Okawa},
  \citenamefont {Lopes},\ and\ \citenamefont {Cardoso}}]{Okawa:2015xma}%
  \BibitemOpen
  \bibfield  {author} {\bibinfo {author} {\bibfnamefont {H.}~\bibnamefont
  {Okawa}}, \bibinfo {author} {\bibfnamefont {J.~C.}\ \bibnamefont {Lopes}}, \
  and\ \bibinfo {author} {\bibfnamefont {V.}~\bibnamefont {Cardoso}},\
  }\href@noop {} {\  (\bibinfo {year} {2015})},\ \Eprint
  {http://arxiv.org/abs/1504.05203} {arXiv:1504.05203 [gr-qc]} \BibitemShut
  {NoStop}%
\bibitem [{\citenamefont {Deppe}\ and\ \citenamefont
  {Frey}(2015)}]{Deppe:2015qsa}%
  \BibitemOpen
  \bibfield  {author} {\bibinfo {author} {\bibfnamefont {N.}~\bibnamefont
  {Deppe}}\ and\ \bibinfo {author} {\bibfnamefont {A.~R.}\ \bibnamefont
  {Frey}},\ }\href@noop {} {\  (\bibinfo {year} {2015})},\ \Eprint
  {http://arxiv.org/abs/1508.02709} {arXiv:1508.02709 [hep-th]} \BibitemShut
  {NoStop}%
\bibitem [{\citenamefont {Balasubramanian}\ \emph {et~al.}(2014)\citenamefont
  {Balasubramanian}, \citenamefont {Buchel}, \citenamefont {Green},
  \citenamefont {Lehner},\ and\ \citenamefont
  {Liebling}}]{Balasubramanian:2014cja}%
  \BibitemOpen
  \bibfield  {author} {\bibinfo {author} {\bibfnamefont {V.}~\bibnamefont
  {Balasubramanian}}, \bibinfo {author} {\bibfnamefont {A.}~\bibnamefont
  {Buchel}}, \bibinfo {author} {\bibfnamefont {S.~R.}\ \bibnamefont {Green}},
  \bibinfo {author} {\bibfnamefont {L.}~\bibnamefont {Lehner}}, \ and\ \bibinfo
  {author} {\bibfnamefont {S.~L.}\ \bibnamefont {Liebling}},\ }\href {\doibase
  10.1103/PhysRevLett.113.071601} {\bibfield  {journal} {\bibinfo  {journal}
  {Phys. Rev. Lett.}\ }\textbf {\bibinfo {volume} {113}},\ \bibinfo {pages}
  {071601} (\bibinfo {year} {2014})},\ \Eprint {http://arxiv.org/abs/1403.6471}
  {arXiv:1403.6471 [hep-th]} \BibitemShut {NoStop}%
\bibitem [{\citenamefont {Craps}\ \emph {et~al.}(2014)\citenamefont {Craps},
  \citenamefont {Evnin},\ and\ \citenamefont {Vanhoof}}]{Craps:2014vaa}%
  \BibitemOpen
  \bibfield  {author} {\bibinfo {author} {\bibfnamefont {B.}~\bibnamefont
  {Craps}}, \bibinfo {author} {\bibfnamefont {O.}~\bibnamefont {Evnin}}, \ and\
  \bibinfo {author} {\bibfnamefont {J.}~\bibnamefont {Vanhoof}},\ }\href
  {\doibase 10.1007/JHEP10(2014)048} {\bibfield  {journal} {\bibinfo  {journal}
  {JHEP}\ }\textbf {\bibinfo {volume} {10}},\ \bibinfo {pages} {48} (\bibinfo
  {year} {2014})},\ \Eprint {http://arxiv.org/abs/1407.6273} {arXiv:1407.6273
  [gr-qc]} \BibitemShut {NoStop}%
\bibitem [{\citenamefont {Craps}\ \emph {et~al.}(2015)\citenamefont {Craps},
  \citenamefont {Evnin},\ and\ \citenamefont {Vanhoof}}]{Craps:2014jwa}%
  \BibitemOpen
  \bibfield  {author} {\bibinfo {author} {\bibfnamefont {B.}~\bibnamefont
  {Craps}}, \bibinfo {author} {\bibfnamefont {O.}~\bibnamefont {Evnin}}, \ and\
  \bibinfo {author} {\bibfnamefont {J.}~\bibnamefont {Vanhoof}},\ }\href
  {\doibase 10.1007/JHEP01(2015)108} {\bibfield  {journal} {\bibinfo  {journal}
  {JHEP}\ }\textbf {\bibinfo {volume} {01}},\ \bibinfo {pages} {108} (\bibinfo
  {year} {2015})},\ \Eprint {http://arxiv.org/abs/1412.3249} {arXiv:1412.3249
  [gr-qc]} \BibitemShut {NoStop}%
\bibitem [{\citenamefont {Deppe}\ \emph {et~al.}(2015)\citenamefont {Deppe},
  \citenamefont {Kolly}, \citenamefont {Frey},\ and\ \citenamefont
  {Kunstatter}}]{Deppe:2014oua}%
  \BibitemOpen
  \bibfield  {author} {\bibinfo {author} {\bibfnamefont {N.}~\bibnamefont
  {Deppe}}, \bibinfo {author} {\bibfnamefont {A.}~\bibnamefont {Kolly}},
  \bibinfo {author} {\bibfnamefont {A.}~\bibnamefont {Frey}}, \ and\ \bibinfo
  {author} {\bibfnamefont {G.}~\bibnamefont {Kunstatter}},\ }\href {\doibase
  10.1103/PhysRevLett.114.071102} {\bibfield  {journal} {\bibinfo  {journal}
  {Phys. Rev. Lett.}\ }\textbf {\bibinfo {volume} {114}},\ \bibinfo {pages}
  {071102} (\bibinfo {year} {2015})},\ \Eprint {http://arxiv.org/abs/1410.1869}
  {arXiv:1410.1869 [hep-th]} \BibitemShut {NoStop}%
\bibitem [{\citenamefont {Okawa}\ \emph {et~al.}(2014)\citenamefont {Okawa},
  \citenamefont {Cardoso},\ and\ \citenamefont {Pani}}]{Okawa:2013jba}%
  \BibitemOpen
  \bibfield  {author} {\bibinfo {author} {\bibfnamefont {H.}~\bibnamefont
  {Okawa}}, \bibinfo {author} {\bibfnamefont {V.}~\bibnamefont {Cardoso}}, \
  and\ \bibinfo {author} {\bibfnamefont {P.}~\bibnamefont {Pani}},\ }\href
  {\doibase 10.1103/PhysRevD.89.041502} {\bibfield  {journal} {\bibinfo
  {journal} {Phys. Rev.}\ }\textbf {\bibinfo {volume} {D89}},\ \bibinfo {pages}
  {041502} (\bibinfo {year} {2014})},\ \Eprint {http://arxiv.org/abs/1311.1235}
  {arXiv:1311.1235 [gr-qc]} \BibitemShut {NoStop}%
\bibitem [{\citenamefont {Basu}\ \emph {et~al.}(2015)\citenamefont {Basu},
  \citenamefont {Krishnan},\ and\ \citenamefont {Saurabh}}]{Basu:2014sia}%
  \BibitemOpen
  \bibfield  {author} {\bibinfo {author} {\bibfnamefont {P.}~\bibnamefont
  {Basu}}, \bibinfo {author} {\bibfnamefont {C.}~\bibnamefont {Krishnan}}, \
  and\ \bibinfo {author} {\bibfnamefont {A.}~\bibnamefont {Saurabh}},\ }\href
  {\doibase 10.1142/S0217751X15501286} {\bibfield  {journal} {\bibinfo
  {journal} {Int. J. Mod. Phys.}\ }\textbf {\bibinfo {volume} {A30}},\ \bibinfo
  {pages} {1550128} (\bibinfo {year} {2015})},\ \Eprint
  {http://arxiv.org/abs/1408.0624} {arXiv:1408.0624 [hep-th]} \BibitemShut
  {NoStop}%
\bibitem [{\citenamefont {Christodoulou}(1999)}]{Christodoulou}%
  \BibitemOpen
  \bibfield  {author} {\bibinfo {author} {\bibfnamefont {D.}~\bibnamefont
  {Christodoulou}},\ }\href {http://www.jstor.org/stable/121023} {\bibfield
  {journal} {\bibinfo  {journal} {Annals of Mathematics}\ }\bibinfo {series}
  {Second Series},\ \textbf {\bibinfo {volume} {149}},\ \bibinfo {pages} {183}
  (\bibinfo {year} {1999})}\BibitemShut {NoStop}%
\end{thebibliography}%

\end{document}